\documentstyle[pra,aps,epsfig]{revtex}

\begin{document}

\draft

\title{Nucleon-Nucleon Correlations and Two-Nucleon Currents in Exclusive 
($e,e'NN$) Reactions}

\author{D.\ Kn\"odler and H.\ M\"uther}

\address{Institut f\"ur Theoretische Physik, Universit\"at T\"ubingen,\\
Auf der Morgenstelle 14, D-72076 T\"ubingen, Germany  }

\author{P.\ Czerski}

\address{Instytut Fizyki Jadrowej, Pl-31-342 Krakow, Poland} 

\date{\today}

\maketitle

\begin{abstract}

The contributions of short-range nucleon-nucleon (NN) correlations, various
meson exchange current (MEC) terms and the influence of $\Delta$ isobar
excitations  (isobaric currents, IC) on exclusive two-nucleon knockout
reactions induced by electron scattering are investigated. The nuclear
structure functions are  evaluated for nuclear matter. Realistic NN
interactions derived in the framework of One-Boson-Exchange model are employed
to evaluate the effects of correlations and MEC in a consistent way. The
correlations correlations are determined by  solving the Bethe-Goldstone
equation. This yields  significant contributions to the structure functions
$W_L$ and $W_T$ of the $(e,e'pn)$ and $(e,e'pp)$  reactions. These
contributions compete with MEC corrections originating  from the $\pi$ and
$\rho$ exchange terms of the same interaction. Special attention is paid to the
so-called 'super parallel'  kinematics at momentum transfers which can be
measured e.g.~at MAMI in Mainz.

\end{abstract}


\section{Introduction}

It is well known that nuclei form a many-body system of interacting nucleons
which exhibit strong correlations beyond the mean field or Hartree-Fock
approximation.  The strong short-range and tensor components of a realistic
model for the nucleon-nucleon (NN) interaction induce corresponding two-body
correlations. Many attempts have been made to explore the details of these
correlations. Photoinduced exclusive two-nucleon knockout experiments like
($\gamma,NN)$ or ($e,e'NN$) reactions seem to be an ideal tool for such
investigations as the cross section is related to the probability that the real
or virtual photon is absorbed by a pair of interacting nucleons. The analysis
of exclusive two-nucleon knockout experiments is restricted to such cases, in
which momentum and energy conservation ensures that the residual nucleus with
$A-2$ nucleons is produced in its groundstate or another well defined bound
state. Due to the progress in accelerator and detector technology such triple
coincidence experiments with good resolution have become possible and first
results have been reported in the literature\cite{blom,onder,rosner}.

In studying such reactions, however, it is important to keep in mind that
there exist various competing mechanisms which all contribute to the
cross section of two-nucleon knockout. Beside the contribution which is
due to the pair correlations in the ground state wavefunction one must
consider the effects of the two-body terms in the operator for the
electromagnetic current. These two-body terms include meson exchange
currents (MEC) which can be derived from the commutator of the charge
density with the nuclear Hamiltonian to obey the continuity equation.
These meson exchange current contributions should be evaluated in a way
consistent with the NN interaction used to describe the nuclear
structure\cite{mec1,mec2}. In the present work we investigate the MEC effects
which are due to the exchange of $\pi$ and $\rho$ mesons. 

In addition, there are contributions to the two-body current which are of
a different origin. Here we, mention the contribution due to the
excitation of intermediate $\Delta$ isobar excitations\cite{ic}. The
contribution of this isobar current (IC) is not constrained by a
continuity equation like the MEC and therefore depends on parameters like
the meson-nucleon $\Delta$ coupling constants and the propagator of the
$\Delta$ in the nuclear medium. The latter be chosen within certain
limits.

Furthermore one should keep in mind that the so-called realistic NN interactions
which we will employ in our study, like the Bonn potential\cite{machleidt}, have
been fitted to the NN scattering phase shifts at energies below the threshold
for $\pi$ productions. Therefore one should also restrict the analysis of
($e,e'NN$) reactions to this regime.

The effects of two-body correlations have often been taken into account by
assuming a local, state-independent correlation function between the
interacting nucleons. From studies of single-nucleon knockout experiments,
however, it is known that the energy dependence of the single-particle
spectral function is essential for the understanding of the reaction
mechanism\cite{eep1,eep2}. Therefore, one would like to consider
dynamical, state-dependent correlations in the investigation of the
two-body knockout, too. One way of achieving this goal is to evaluate the
matrix elements for the photon absorption by a correlated pair of nucleons
directly in terms of the Brueckner $G$ matrix and the corresponding
propagators for the nucleons.

In this manuscript we will describe a technique that allows to evaluate
the nuclear matrix elements for the absorption of a photon by a correlated
pair of nucleons, the MEC and the IC in a consistent scheme which includes
the effects of the final state interaction of the outgoing nucleons with
the residual nucleus in a mean field approximation. As a first step, this
technique is applied to the absorption of the photons by nucleons in
infinite nuclear matter. This allows to compare the relative importance of
these different mechanisms and their interference under various
kinematical conditions.

It is of course a serious disadvantage of such a study in nuclear matter
that it does not provide results for a cross section which can directly be
compared with experimental data produced for a specific target nucleus. In
particular, it is impossible to take advantage of the fact that reactions
leading to specific final states of the residual nucleus can be selective
for one of the two-nucleon knockout mechanisms discussed above. Such a
feature has been observed in theoretical studies of ($e,e'pp$) an ($e,e'pn$)
reactions on $^{16}{\rm O}$\cite{carlot,carlopn}. 
On the other hand, however, a study of nuclear matter shall exhibit 
general features which are independent on the
specific target nucleus considered and the corresponding long-range or 
low-energy correlations. A comparison of results for nuclear
matter with those obtained for finite systems should enable us to
disentangle effects which are global and due to the short-range behaviour
of two nucleon correlations from those sensitive to surface and finite
size effects. Furthermore such a study shall also allow to explore
systematically the importance of e.g.~the MEC originating from the $\rho$ meson,
which is often ignored. Finally, a study in nuclear matter shall also help to 
choose efficient setups for experiments which are selective in the sense that 
they  yield results which are particularly sensitive to either the exploration 
of short-range correlations, MEC or IC contributions to the two-nucleon
knockout reaction.

After this introduction we will present a description of the methods used
to calculate the various contributions to the photoinduced two-nucleon
knockout reactions in section 2. The detailed discussion of results is
presented in section 3. Some concluding remarks are added in the final
section 4.

\section{The Cross Section and the Nuclear Current}

We start from the definition of the nine-fold differential cross section
of the $(e,e'2N)$ reaction \cite{wq1,wq2} 
\begin{eqnarray}
\frac{{\rm d}^9\sigma}{{\rm d}\tilde{E_1}{\rm d}\tilde{\Omega_1} {\rm d}
\tilde{E_2}{\rm d}\tilde{\Omega_2} {\rm d}E_e'{\rm d}\Omega_e'} 
&= &\frac{1}{4}\,\frac{1}{(2\pi)^9}\, \tilde{p_1}\, \tilde{p_2}\, \tilde{E_1}\, 
\tilde{E_2}\,
\sigma_{\rm Mott} \,
\Big\{ v_C W_L + v_T W_T +v_S W_{TT} + v_I W_{LT} \Big\}\, 
\nonumber \\ 
&& \times (2\pi)\, \delta (E_f-E_i)
\end{eqnarray}
where $\tilde{E_1},\tilde{E_2}$ and $\tilde{p_1},\tilde{p_2}$ denote the 
energies and momenta of the outgoing nucleons, respectively.  The
virtual photon created in the electron scattering process carries momentum
$\vec{q}$ and energy $\omega$. The leptonic structure functions 
$v_i$ ($i=C,T,S,I$) are defined by
\begin{eqnarray}
v_C&=& \Big( \frac{q_{\mu}q^{\mu}}{\vec{q}\,^2} \Big)^2 \nonumber \\
v_T&=& \tan^2 \frac{\theta_e}{2} - \frac{1}{2} \Big(
\frac{q_{\mu}q^{\mu}}{\vec{q}^{\,2}} \Big) \nonumber \\
v_I&=& \frac{q_{\mu}q^{\mu}}{\sqrt{2}|\vec{q}\,|^3}\,(E_e+E_e')\,
\tan\frac{\theta_e}{2}\nonumber \\
v_S&=& \frac{q_{\mu}q^{\mu}}{2\vec{q}\,^2}
\end{eqnarray}
Here, $\theta_e$ is the angle of the scattered electron with respect to the 
incident electron beam and $E_e,E_e'$
are the energies of the incident and the scattered electron, respectively.

The nuclear structure functions $W_i$ ($i=L,T,TT,LT$) contain the matrix
elements of the nuclear current operator for a given photon polarization
$\lambda$. These matrix elements are calculated for the current operator
$\vec{J}=\vec{J}^{(1)}+\vec{J}^{(2)}$ which accounts for the different
processes contributing to either the absorption of the photon energy and
momentum on a single nucleon [cf.\ Figure \ref{fig2}, diagrams (a) and
(b)] or on a pair of nucleons [cf.\ Figure \ref{fig2}(c) and the diagrams
in Figures \ref{fig3} and \ref{fig4}]. As a main aim of the investigation
of the two-nucleon knockout reaction is to disentangle these different
contributions all mechanisms have to be described with a consistent set of
operators in order to avoid ambiguities concerning e.g. coupling constants
or cutoff masses. Within the framework of nuclear matter where the
nucleons are described by plain-wave states we perform calculations for
the different components of the nuclear current operator presented in
detail in the following sections.

\subsection{Ground-State Correlations and the Nuclear $G$ Matrix}

\begin{center}
\begin{figure}[h]
\epsfig{file=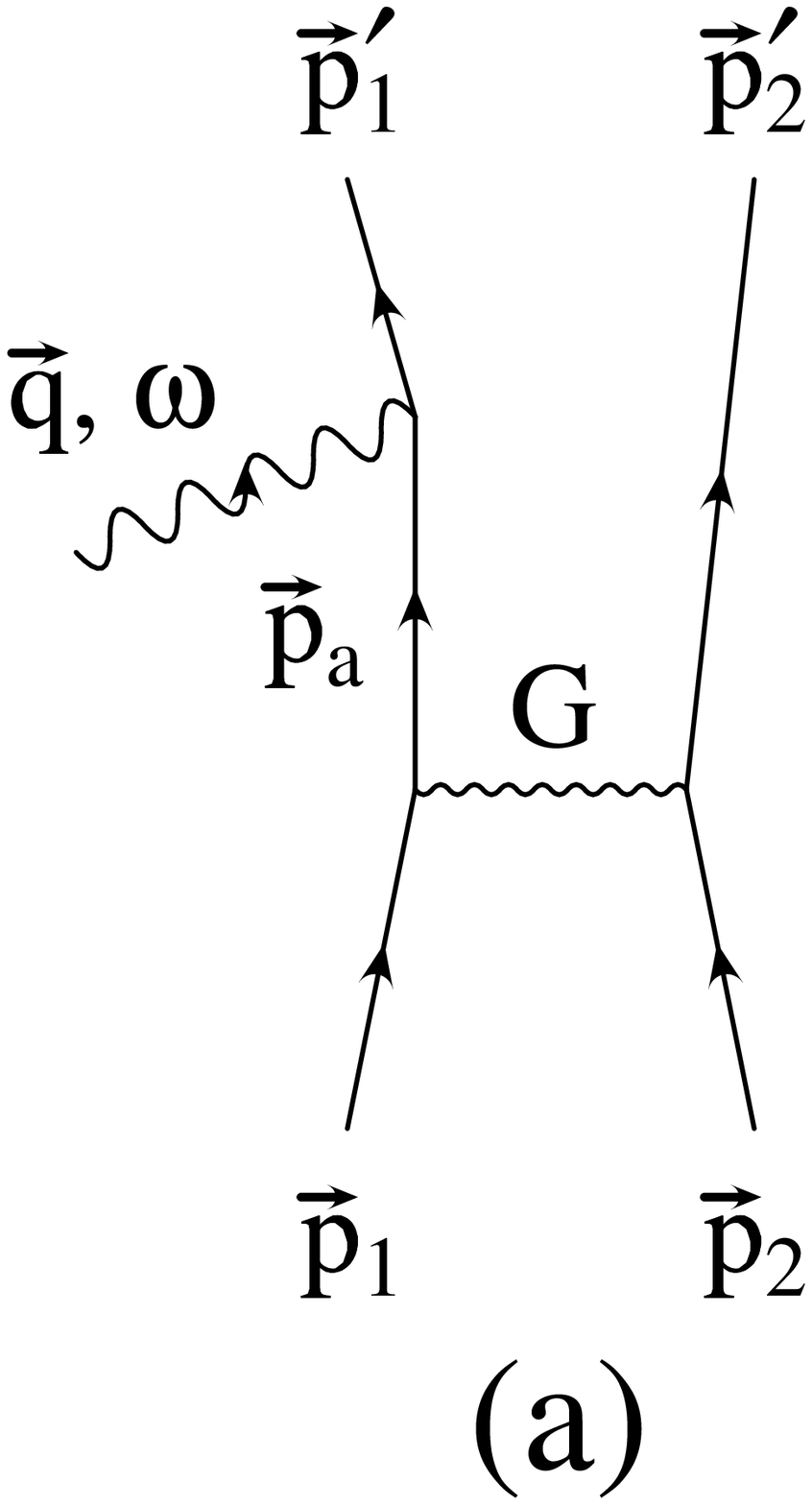,scale=0.35}\hspace{1cm}
\epsfig{file=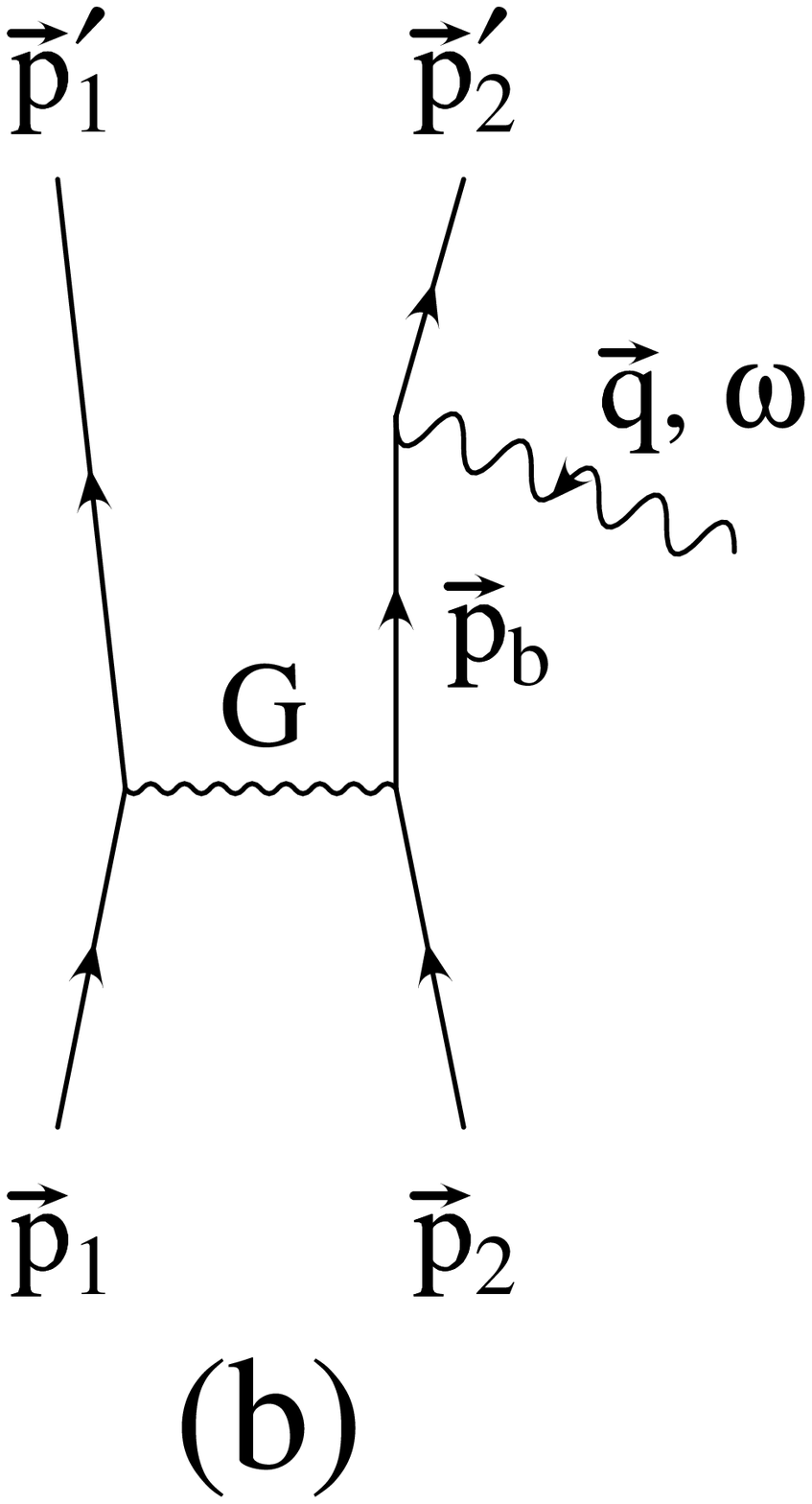,scale=0.35}\hspace{.5cm}
\epsfig{file=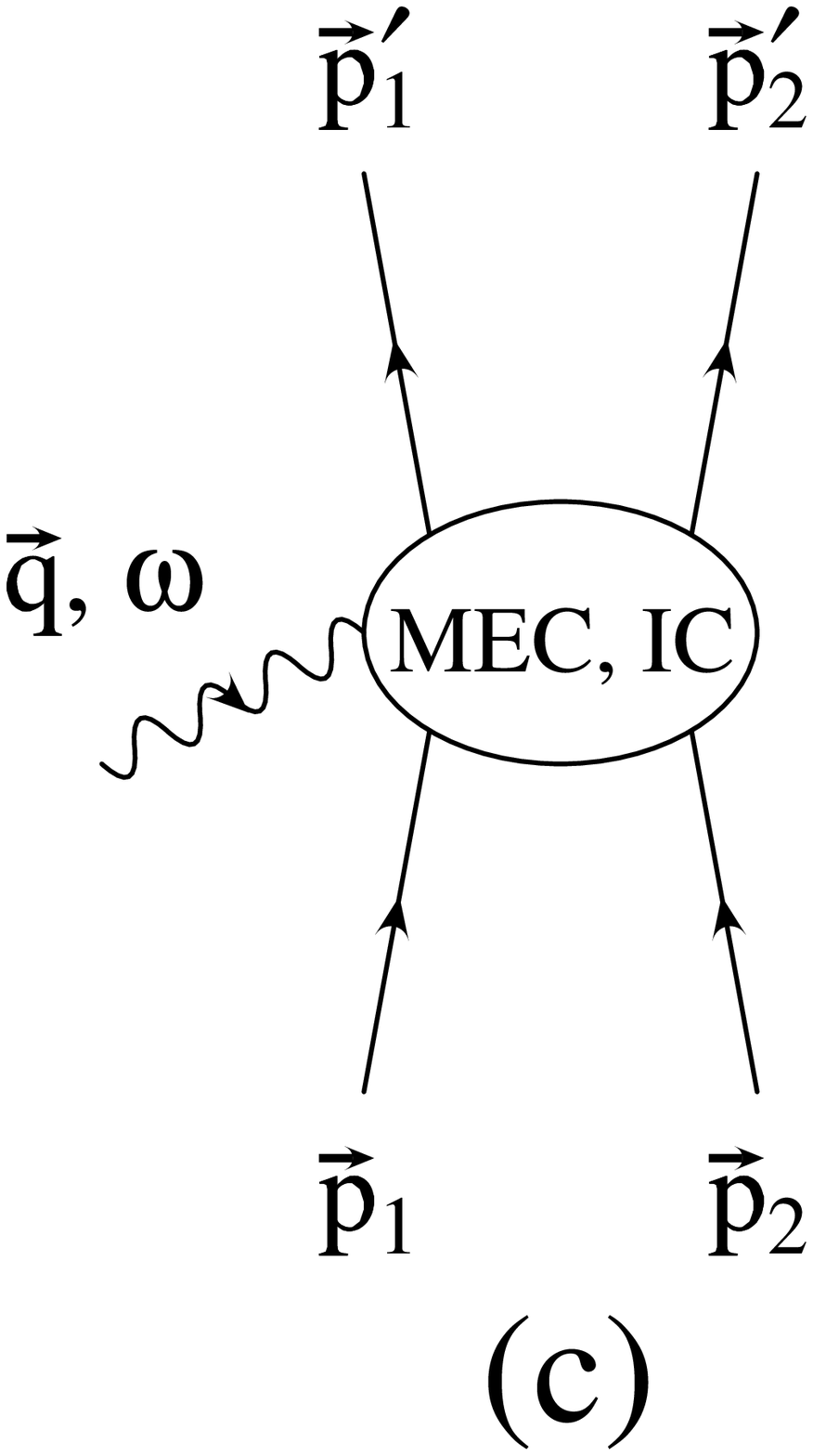,scale=0.35}
\vspace{.5cm}
\caption{\label{fig2}
Diagrams for the different processes contributing to the $(e,e'2N)$
reaction. Diagram (a) and (b) show the absorption of the photon by a
single nucleon. The nucleon-nucleon correlations are described by the $G$
matrix. Diagram (c) depicts photon absorption via meson exchange (MEC) or
isobaric currents (IC) (cf.\ Figures \ref{fig3} and \ref{fig4}).}
\end{figure}
\end{center}

Figure \ref{fig2} shows in diagram (a) and (b) the absorption of a
photon carrying momentum $\vec{q}$ and energy $\omega$ on a single
nucleon. This single nucleon is part of a correlated nucleon-nucleon pair.
Here, correlations arising from the short-range and tensor components of
the nucleon-nucleon interaction are calculated in terms of
the nuclear $G$ matrix \cite{correlations}.
As an example, we will discuss the evaluation of the diagram displayed in
part (a) of Figure \ref{fig2} using the notation displayed there.
The matrix element which is required for the evaluation of the nuclear
structure functions is calculated in a basis of plane-wave states 
\begin{equation}
\int{\rm d}\vec{p}_a\,
\langle \vec{p}_1{'} \,|\, H_{\gamma NN} \,|\, \vec{p}_a \rangle\,
S(p_a,p_2')\,
\langle
\vec{p}_a\, \vec{p}_2{'} \, |\, G \,|\, \vec{p}_1\, \vec{p}_2 \rangle.
\label{eq:21}
\end{equation}
Here and in the following, $\vec{p}_i$ denote the momenta of two
uncorrelated 
nucleons in the target system. The center of mass momentum of this pair $\vec P
= \vec{p}_1 + \vec{p}_2$ defines the total missing momentum of the two-nucleon
knockout process. The single-particle energies referring to these momenta
$\varepsilon_i= \varepsilon (p_i)$  will be parametrized in a simple 
effective mass approximation
\begin{equation}
\varepsilon(p) = \frac{p^2}{2m^*} + U \label{eq:effm}
\end{equation}
with an effective mass $m^*$ and a potential $U$ adjusted in such a way that
(\ref{eq:effm}) yields a good approximation for the single-particle
energies
evaluated in a Brueckner-Hartree-Fock calculation of nuclear
matter\cite{correlations}. The sum of the single-particle energies referring to
the momenta $\vec{p}_i$ defines the total removal or missing energy  
$\varepsilon(p_1) + \varepsilon(p_2)$. The momenta $\vec{p}_1{'}$ and 
$\vec{p}_2{'}$
define the momenta of the outgoing nucleons inside nuclear matter. The
directions of these momenta correspond to the direction of the momenta
$\tilde{p_i}$ finally observed in the detector. The modulus of these vectors
$\vec{p}_i'$ is determined in such a way that the single-particle energies
of the
nucleons inside nuclear matter, calculated according to (\ref{eq:effm})
coincide 
with the free energies of the nucleons moving with momenta
$\tilde{\vec{p}}_i$. In
this way, we account for the retardation of the outgoing nucleons in the
mean
field of the remaining nucleus. The energy conservation can be expressed by
\begin{equation}
\varepsilon(p_1)+\varepsilon(p_2)+\omega=\varepsilon(p_1{'})+
\varepsilon(p_2{'}) =
\frac{\tilde{p_1}^2}{2m_N} + \frac{\tilde{p_2}^2}{2m_N}\,, \label{eq:econ}
\end{equation}
while the conservation of total momentum  yields
\begin{equation}
\vec{p}_1+\vec{p}_2+\vec{q}=\vec{p}_a+\vec{p}_2{'}+\vec{q} =
\vec{p}_1{'}+\vec{p}_2{'}\,.\label{eq:momcon}
\end{equation} 
The momentum $\vec{p}_a$ in (\ref{eq:21}) and (\ref{eq:momcon}) denotes
the 
momentum of the intermediate single-nucleon state before the absorption of 
the photon (see also part (a) of Figure \ref{fig2}). The propagation of the
intermediate two-nucleon state is described in terms of the propagator
\begin{equation}
S(p_a,p_2') = \frac{{\cal Q}(p_a,p_2')}{\varepsilon(p_1)+\varepsilon(p_2)
-\varepsilon(p_a)-\varepsilon(p_2') + {\rm i}\eta}
\end{equation}
with a Pauli operator ${\cal Q}(p_a,p_2')$ which ensures that the nucleons are
propagating in states above the Fermi sea.

The absorption of the real or virtual photon carrying
momentum $\vec{q}$ by the single (off-shell) nucleon with initial momentum
$\vec{p}_a$ and
final momentum $\vec{p}_1{'}$ is described by the Hamiltonian
\begin{equation}
H_{\gamma NN}=i\,e\,\frac{\mu_N}{2m_N} (\vec{\sigma}\times\vec{q}\,)
\cdot \vec{\varepsilon} +e \,\frac{\delta_N}{2m_N}
(\vec{p}_a+\vec{p}_1{'})
\cdot \vec{\varepsilon}
\end{equation}
where $\delta_N=1$ for a proton and $\delta_N=0$ for a neutron
\cite{photon}. The vector $\vec{\varepsilon}$ refers to the polarization of the
absorbed photon. Note that we employ here the non-relativistic reduction of the
one-body current ignoring possible off-shell effects, which are due to a
possible medium modification of the nucleons in the nuclear medium\cite{boffbu}.  

As the final ingredient, the nuclear $G$ matrix has to be calculated. For
a realistic nucleon-nucleon interaction described by a potential $V$, the
Bethe-Goldstone equation
\begin{equation}
G=V+V\frac{{\cal Q}}{W-H_0+{\rm i}\eta}G
\end{equation}
has to be solved. Again the operator ${\cal Q}$ ensures that the
Pauli principle is obeyed in the nuclear medium. The starting energy $W$ is
the sum of the single-particle energies $\epsilon(p_1) + \epsilon(p_2)$.
The nuclear $G$ matrix
conserves the center of mass momentum of the interacting pair of nucleons
$\vec{p}_1+\vec{p}_2=\vec{p}_a+\vec{p}_2{'}$ but
leads to a redistribution of relative momenta between the
two nucleons going up to several hundred MeV/c. The calculation of these $G$
matrix elements is performed in a partial wave basis using the BONN A
potential \cite{machleidt}. Also the effective masses $m^*$ and values
of the single-particle potential $U$ are determined for this potential.

\subsection{Meson Exchange Currents}

Meson exchange currents (MEC) represent another possibility of absorbing a
photon on a pair of nucleons. In this case, the real or virtual photon is
absorbed in coincidence with the exchange of a virtual meson ($\pi$,
$\rho$, $\sigma$, $\omega$, $\ldots$) between the nucleon-nucleon pair.
The corresponding current operators can be derived either by minimal
coupling or via the continuity equation using meson exchange potentials
\cite{mec1,mec2}. 

\begin{figure}[h]
\epsfig{file=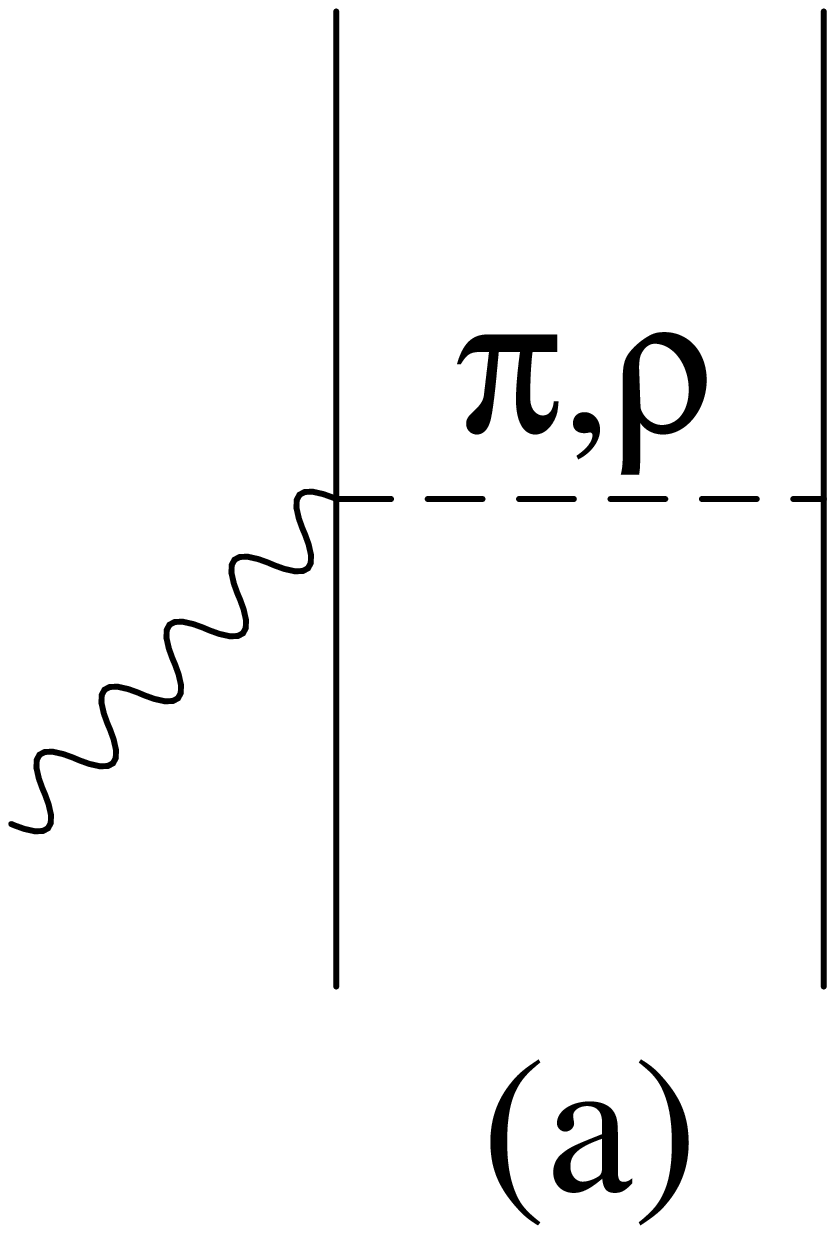,scale=0.3}\hspace{1.8cm}
\epsfig{file=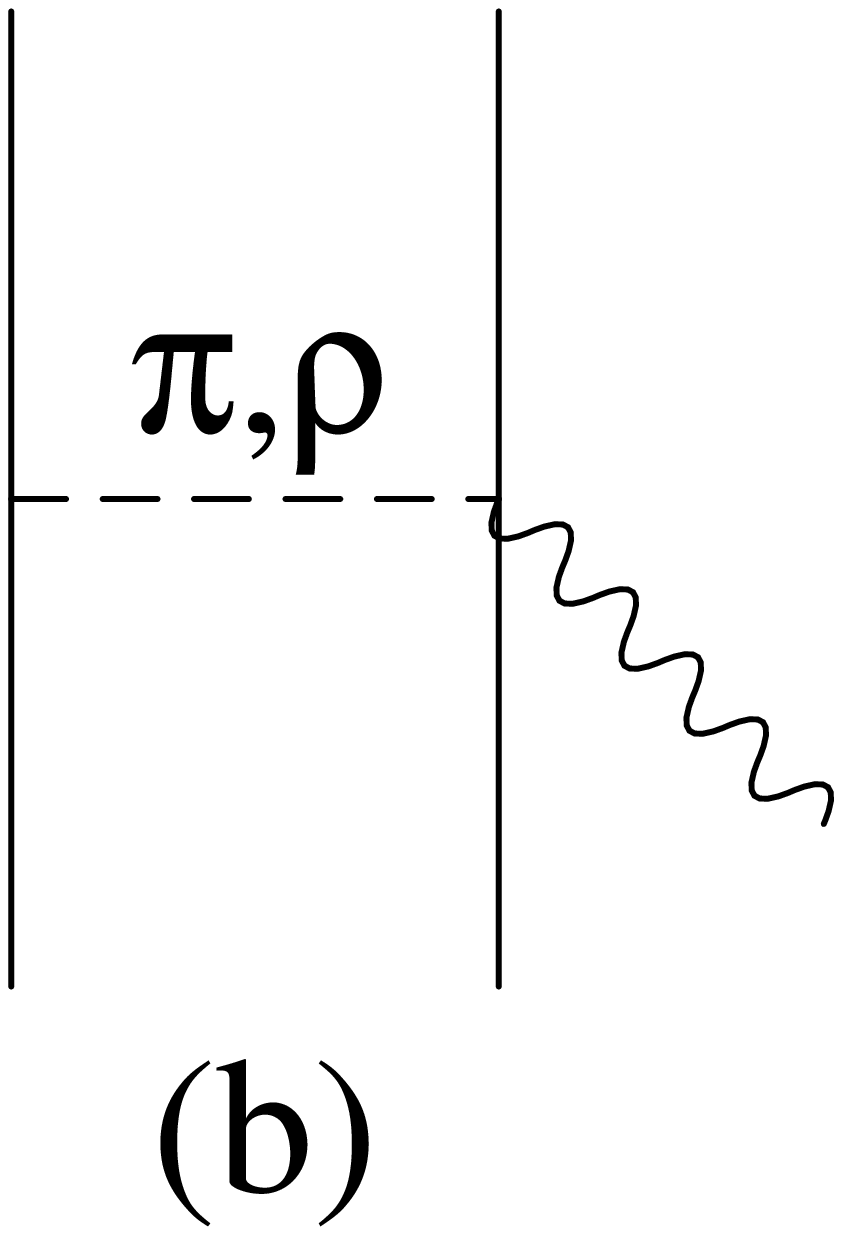,scale=0.3}\hspace{1.5cm}
\epsfig{file=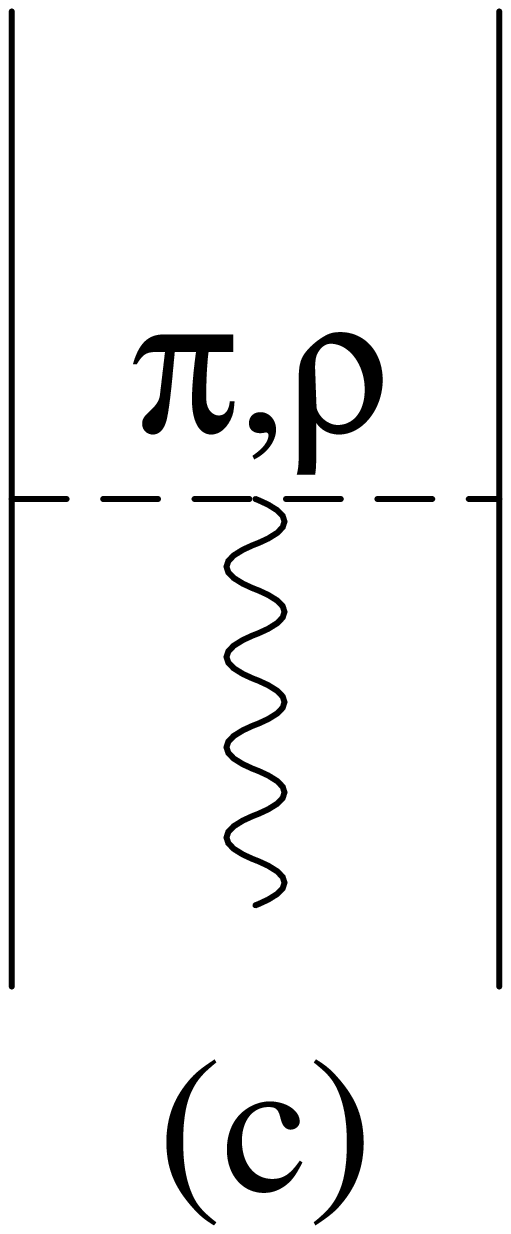,scale=0.3}\hspace{1.8cm}
\epsfig{file=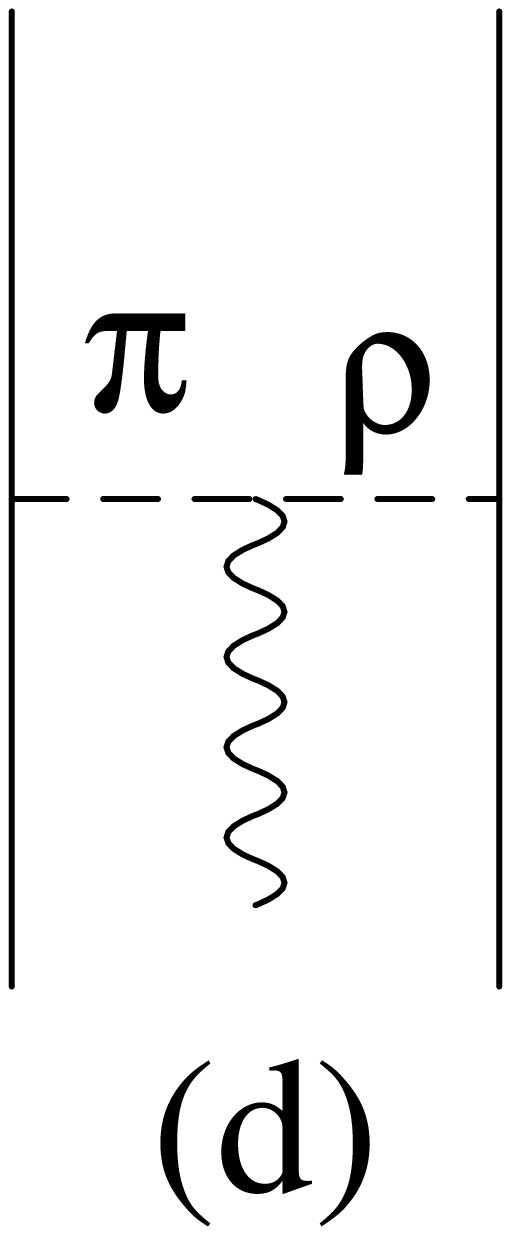,scale=0.3}
\vspace{.5cm}
\caption{\label{fig3}
MEC operators: Diagrams (a) and (b) show the seagull term for the exchange
of a $\pi$ or a $\rho$ meson respectively, diagram (c) shows the
meson-in-flight contribution, and (d) illustrates the coupling of the
photon to a $\pi$ and a $\rho$ meson.}
\end{figure}

Starting with a single-particle charge density $\rho=\frac{1}{2} \,e
\sum_{i=1}^A (1+\tau_{i,z}) \,\delta(\vec{r}-\vec{r}_i)$ and inserting as an
example the non-relativistic reduction of the
$\pi$-exchange potential
\begin{equation}
V_{\pi\!N\!N}=-v_{\pi}(k) \,(\vec{\sigma}_1\cdot\vec{k})\,
(\vec{\sigma}_2\cdot\vec{k}) \,(\vec{\tau}_1\cdot \vec{\tau}_2)
\end{equation}
with $v_{\pi}(k) = \frac{f_{\pi N N}^2}{m_{\pi}^2}
\,\frac{1}{m_{\pi}^2+k^2}$ in the continuity equation
\begin{equation}
\vec{\nabla}\cdot\vec{j}+ i\,\big[\,H\,,\,\rho\,\big]=0
\end{equation}
yields the corresponding MEC operator
\begin{eqnarray}\label{mecop}
\vec{j}_{\pi}(\vec{k}_1,\vec{k}_2)&=&-i\,e\,\frac{f_{\pi
N N}^2}{m_{\pi}^2}\,(\vec{\tau}_1\times
\vec{\tau}_2)_z \nonumber\\
&&\times\Bigg[   
\frac{\vec{\sigma}_1\,(\vec{\sigma}_2 \cdot\vec{k}_2)}{m_{\pi}^2+k_2^2}
-\frac{\vec{\sigma}_2 \,(\vec{\sigma}_1\cdot\vec{k}_1)}{m_{\pi}^2+k_1^2}
-\frac{(\vec{\sigma}_1\cdot\vec{k}_1)(\vec{\sigma}_2\cdot\vec{k}_2)}
   {(m_{\pi}^2+k_1^2)(m_{\pi}^2+k_2^2)}(\vec{k}_1-\vec{k}_2)
\Bigg],
\end{eqnarray}
which can be split into the so-called seagull [first two terms, cf.\ diagrams 
(a) and (b) of Figure \ref{fig3}] and pion-in-flight [cf.\ diagram (c) of Figure
\ref{fig3}] terms. The isospin structure of this current operator yields a
non-vanishing contribution only for the exchange of charged pions and 
therefore it will contribute to the knockout of a proton-neutron pair only.

In order to regularize the $\pi$-exchange potential at large momentum transfers
$k$,
the meson-nucleon vertices are multiplied by monopole type form factors
\begin{equation}
F(q)=\frac{\Lambda_{\pi}^2 - m_{\pi}^2}{\Lambda_{\pi}^2 + k^2 }
\end{equation}
This regularization must be taken into account before deriving the 
MEC operators to ensure that the continuity
equation is satisfied up to second order in the nucleon-nucleon
interaction.

In the same manner, the MEC operators corresponding to the exchange of a
$\rho$ meson can be derived\cite{mec1,towner}. For the case of two-nucleon
knock-out it has been observed, however, that the effects of the $\rho$ MEC
corrections are not of particular significance\cite{vander} and could be
simulated the cut-off parameter for the pion\cite{mache}. Therefore we do not
include the effects of $\rho$ MEC in the present investigation.

Another type of MEC operators is usually
quoted as 'model dependent' in the sense that it cannot be deduced using
the continuity equation. These operators either refer to the absorption of
the photon on two different mesons [Figure \ref{fig3}(d)] or to processes
involving the excitation or de-excitation of an intermediate $\Delta$
resonance (depicted in Figure \ref{fig4}).

To calculate the matrix elements of the current operator
$\vec{j}_{\pi}(\vec{k}_1,\vec{k}_2)$, we refer to the kinematical setting
described in the previous section. The momentum transfers $\vec{k}_1$
and $\vec{k}_2$ can then be determined via
\begin{equation}
\vec{k}_1=\vec{p}_1{'}-\vec{p}_1 \;\;\; {\text and} \;\;\;
\vec{k}_2=\vec{p}_2{'}-\vec{p}_2.
\end{equation}
Finally, the matrix element for the exchange of a $\pi$-meson
in a plane-wave basis can be written as
\begin{equation}
\langle \vec{p}_1{'} \, \vec{p}_2{'} \,|\,
\vec{j}_{\pi}(\vec{k}_1,\vec{k}_2) \,|\, \vec{p}_1 \, \vec{p}_2 \rangle.
\end{equation}

\subsection{Isobaric Currents}

\begin{figure}[h]
\epsfig{file=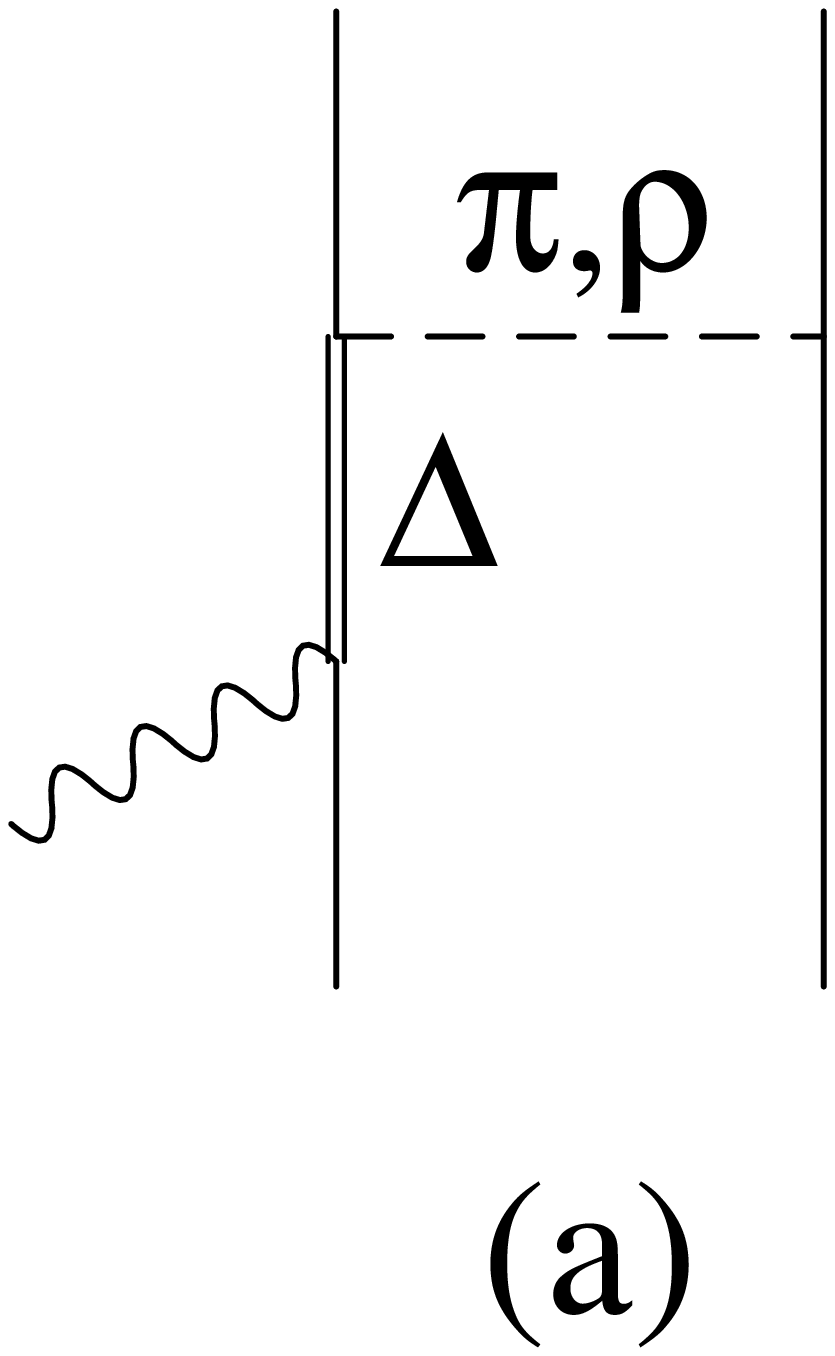,scale=0.3}\hspace{1.8cm}
\epsfig{file=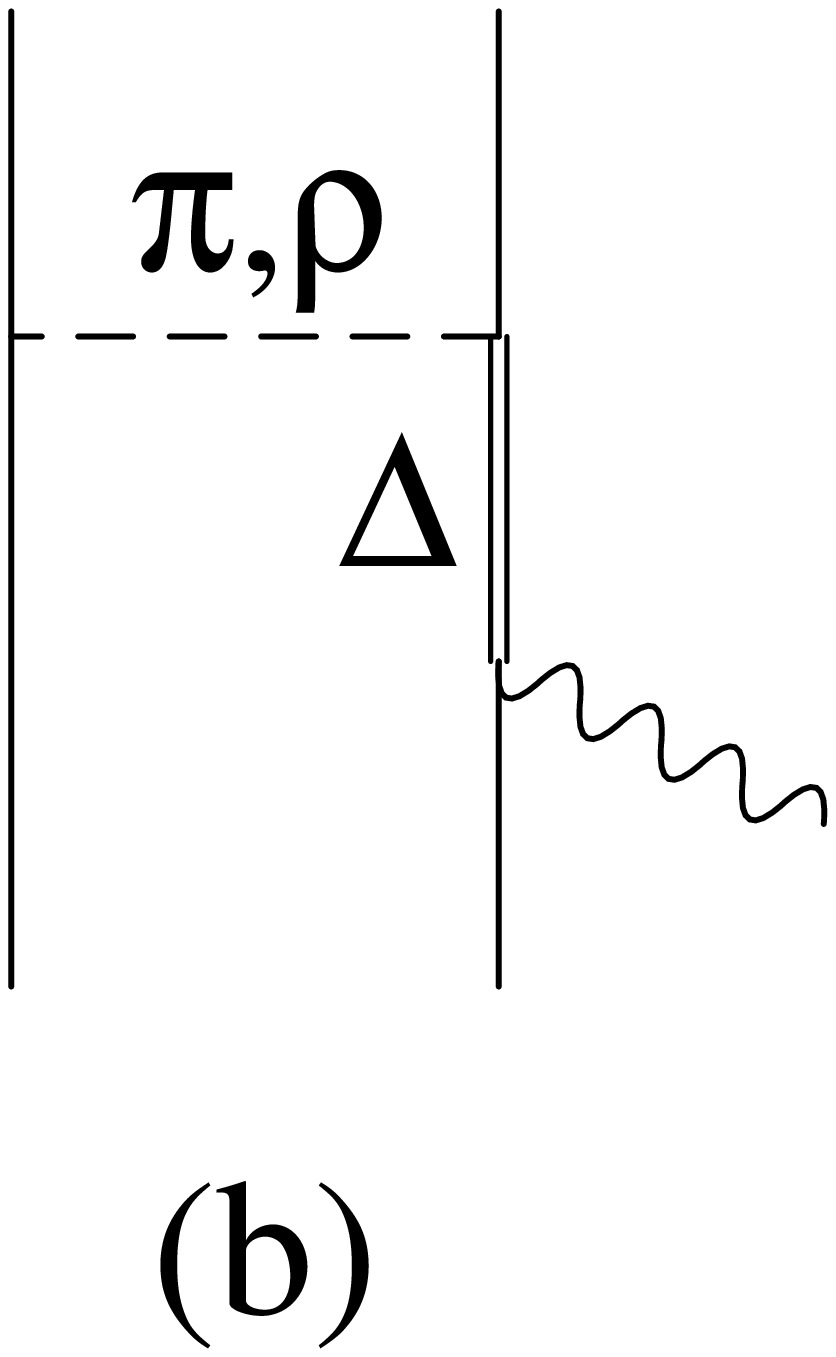,scale=0.3}\hspace{.5cm}
\epsfig{file=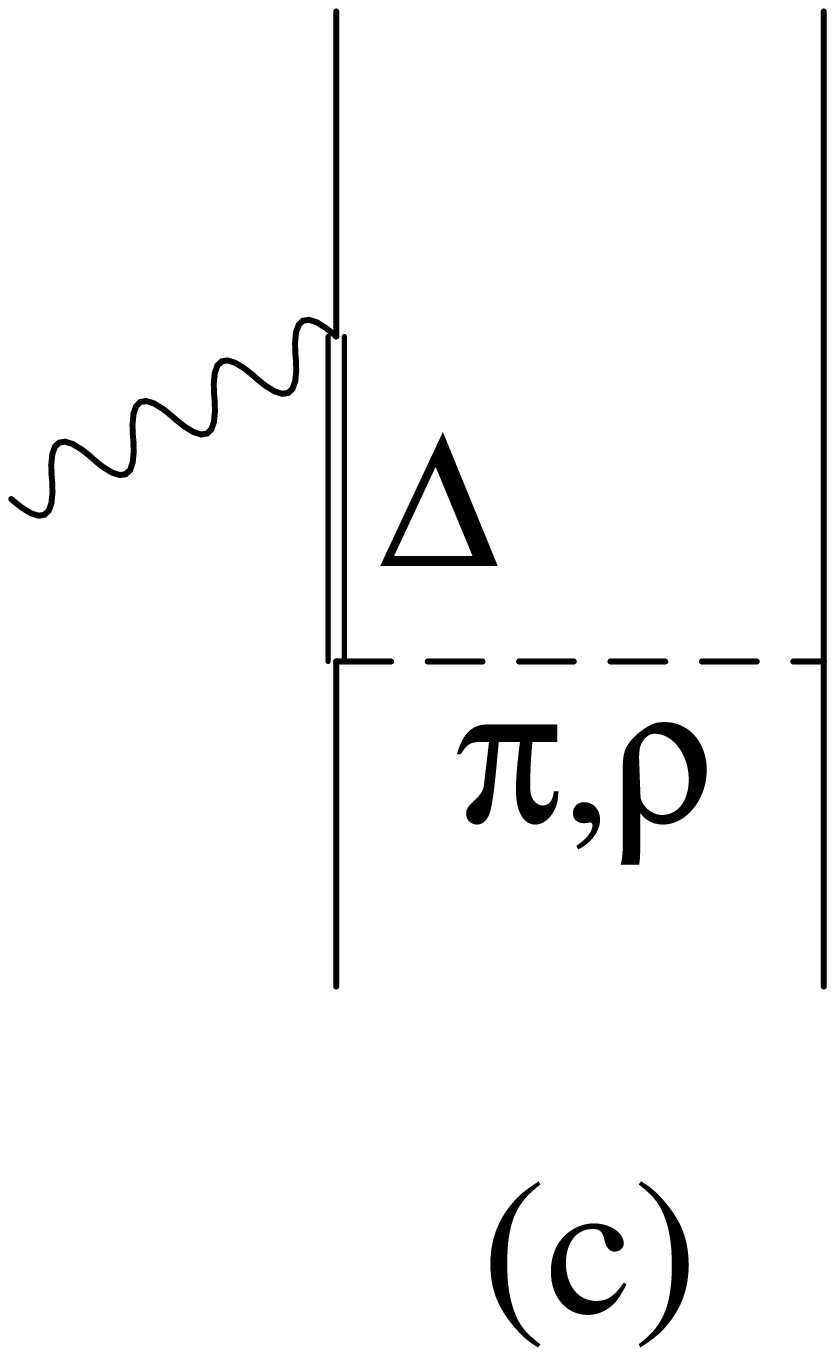,scale=0.3}\hspace{1.8cm}
\epsfig{file=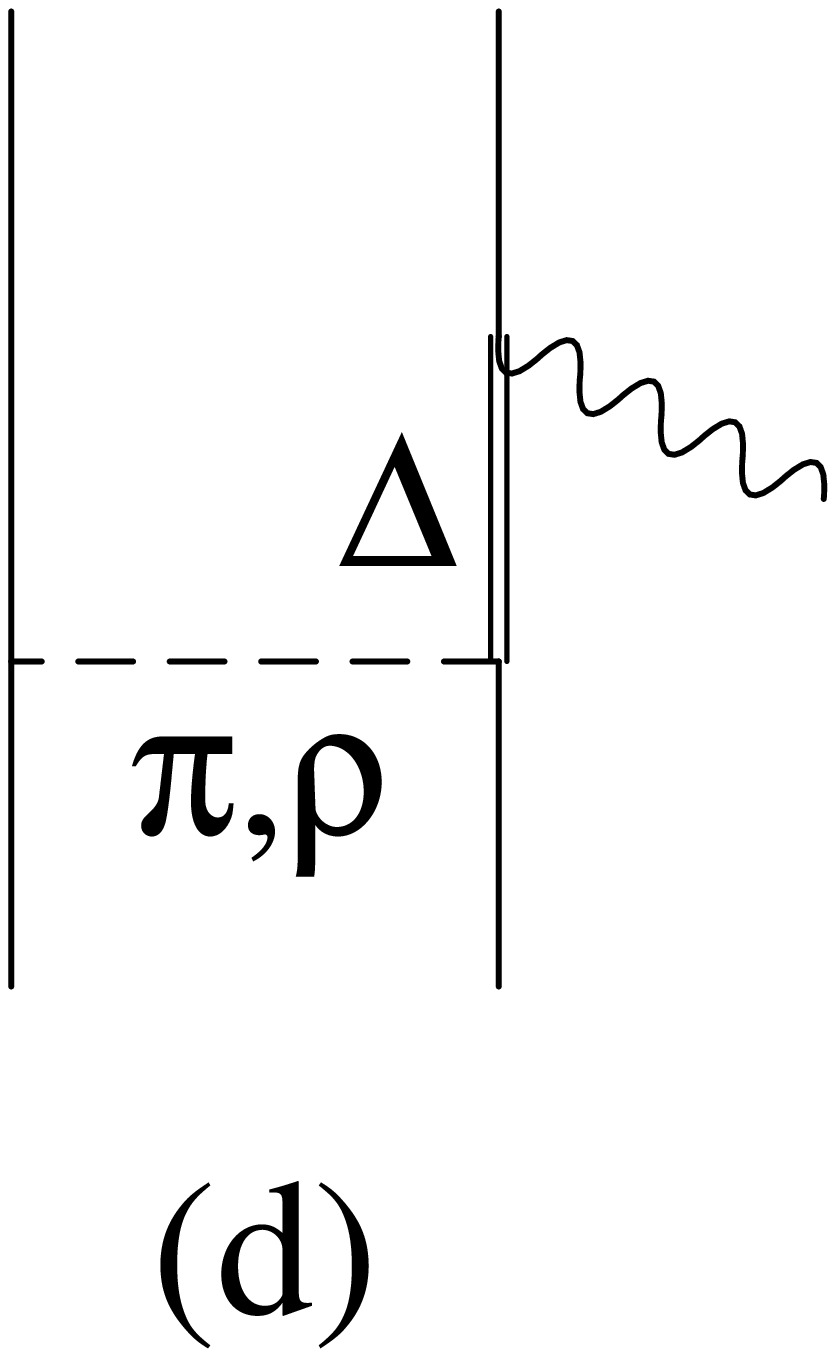,scale=0.3}
\vspace{.5cm}
\caption{\label{fig4}
Isobaric current contributions. Diagrams (a) and (b) show the excitation
of the $\Delta$ resonance due to the absorption of a photon. The
intermediate resonance is de-excited by the exchange of a $\pi$ or a
$\rho$ meson. In diagrams (c) and (d) meson exchange creates a $\Delta$
resonance which is de-excited by the absorption of a photon.}
\end{figure}

Considering the inclusion of an intermediate $\Delta$ resonance leeds to
two different reaction processes. Either the absorbed photon
creates a $\Delta$ resonance which afterwards leads to the exchange of a
charged or uncharged meson [cf.\ Figures \ref{fig4} (a) and (b)] or the
excitation of an intermediate $\Delta$ resonance is caused by the exchange
of a meson and the final photon absorption is responsible for the
de-excitation of the resonance [cf.\ Figure \ref{fig4} (c) and (d)].

The construction of the corresponding operators is not possible via the
insertion of a nucleon-nucleon potential in the continuity equation. In
contrast to the deduction of the MEC operators in the previous section,
here the operators for the isobaric currents are constructed by
simply considering the three different components of the absorption
process.
According to \cite{ic},
diagram (a) in Figure \ref{fig4} may be described by the operator
\begin{equation}
\vec{j}_{\Delta}^{\,(a)}=\,-\frac{{\rm i}}{9}\,
\frac{f_{\pi N N}
f_{\pi N \Delta} f_{\gamma N \Delta}}{m_{\pi}^3}\,
S_{\Delta}^{(a)}\, \frac{\vec{\sigma}_2 \!\cdot\! \vec{k}}{m_{\pi}^2
+ \vec{k}^2} \left\{ 4(\vec{\tau}_2)_z\, \vec{k} + (\vec{\tau}_1
\times \vec{\tau}_2)_z\, (\vec{k} \times \vec{\sigma}_1) \right\} \times
\vec{q},
\end{equation}
which besides the momentum, spin, and isospin structure contains a
propagator $S_{\Delta}$ for the intermediate $\Delta$ resonance.
Transverse by construction, this operator satisfies the continuity
equation and cannot contribute to $(e,e'2N)$ reactions with purely
longitudinally polarized photons. In contrast to the MECs derived using
the continuity equation, isobaric currents also allow for the exchange of
uncharged mesons. They therefore contribute to the knockout of
proton-proton pairs.

Similar to the outlined procedure, the derivation of the current operators
for the three remaining diagrams in Figure \ref{fig4} may be performed.
Of
course, the excitation of a $\Delta$ resonance by the absorption of a
photon [diagrams (a) and (b)] requires a structure of the $\Delta$
propagator different from the propagators for diagrams (c) and (d).  The
calculation of $S_{\Delta}$ is performed according to \cite{ic} considering
the invariant energy of the final two-nucleon state or the binding
energies for the initial states plus the photon energy $\omega$ [see
(\ref{eq:econ})]. Following this procedure, one gets for
diagram (a) of Figure \ref{fig4}
\begin{equation}\label{delprop}
S_{\Delta}=\frac{1}{\varepsilon(p_1) + \omega - \varepsilon_{\Delta}
+ \frac{\rm i}{2} \Gamma_{\Delta}(\omega)}
\end{equation}
where $\varepsilon_{\Delta}= \frac{(\vec{p}_1 + \vec q\,)^2}{2m_{\Delta}} 
+m_{\Delta} -m_N$ 
is the energy of the $\Delta$ state with momentum
$\vec{p}_1 + \vec q$ including the $\Delta N$ mass difference.
The Delta resonance is also given a decay width $\Gamma_{\Delta}
(\omega)$ which shows a variation depending on the energy of the 
absorbed photon. 
For photon energies $\omega$ below 300 MeV we chose an energy dependent width 
according to \cite{oset}
\begin{equation}
\Gamma_{\Delta}(\omega)=\frac{8f_{\pi NN}^2}{12\pi}\, 
\frac{(\omega^2-m_{\pi}^2)^{\frac{3}{2}}}{m_{\pi}^2}\,
\frac{m_{\Delta}-m_N}{\omega}.
\end{equation}
Approximations which have often been used for the $\Delta$ propapgator
include e.g. \cite{ryckdelta}
\begin{equation}\label{propag1}
S_{\Delta}^{approx.}=\frac{1}{m_{\Delta}- m_N-\omega -\frac{\rm
i}{2}\Gamma_{\Delta}(\omega)}
\end{equation}
or even the static propagator
\begin{equation}\label{propag2}
S_{\Delta}^{static}=\frac{1}{m_{\Delta}- m_N}.
\end{equation}

In our calculations all nucleon-meson and $\Delta$-meson vertices
are multiplied by a monopole type form factor with the same cutoff mass
$\Lambda_{\pi}$ as in the previous section.

\section{Results and Discussion}

All calculations are performed in nuclear matter at saturation density
($k_F=1.35\,{\rm fm}^{-1}$). The coupling constants for the $\pi$ exchange
are $f_{\pi N N}=1.005$, $f_{\pi N \Delta}=2f_{\pi N N}$, and $f_{\gamma N
\Delta}=0.12$. We used a cutoff mass for the $\pi$-nucleon form factor of
$\Lambda_{\pi}=1.3\,{\rm GeV}$. In all considered cases, the MEC
contributions contain the seagull and in-flight currents for the exchange
of a $\pi$ meson. Additional MEC contributions like the photon absorption
on a pair of $\pi$ and $\rho$ mesons were found to give negligible
contributions in all kinematical setups we investigated. The $\pi$-nucleon
vertex is compatible with the $\pi$ exchange part of the
One-Boson-Exchange potential BONN A\cite{machleidt} we used to
determine the $G$ matrix for calculating the effects of NN correlations.
The same potential has also been used to determine the single-particle
energies for the nucleons in the nuclear medium which were parametrized
according to (\ref{eq:effm}) by $m^*=623\, {\rm MeV}$ and $U=-86.8\, {\rm
MeV}$.

\begin{figure}[h]
\hspace{1cm}\epsfig{file=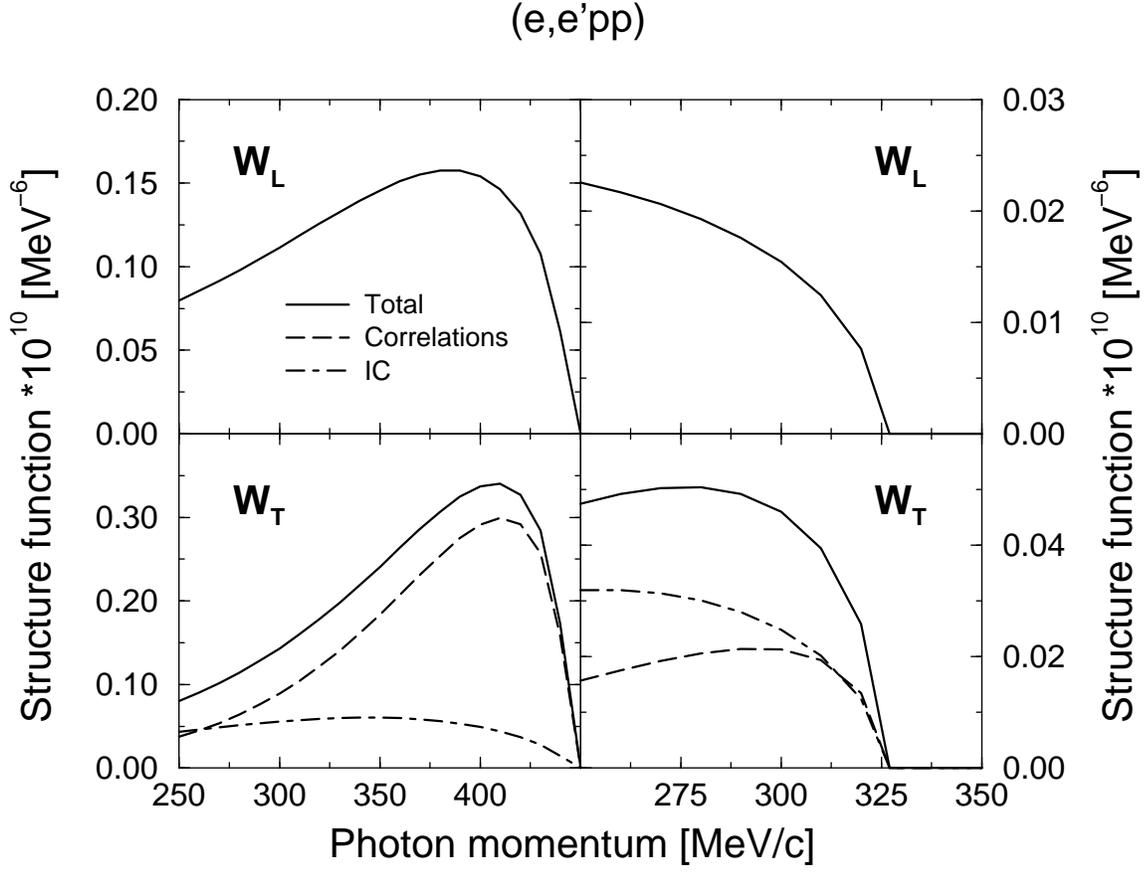,scale=0.7}
\vspace{.5cm}
\caption{\label{fig5}
Longitudinal (upper part) and transverse structure functions (lower part)
for the knockout of a proton-proton pair in a 'super parallel' kinematical
situation with angles $\theta_{p,1}'=0^{\rm o}$ and
$\theta_{p,2}'=180^{\rm o}$ of the two protons with respect to the
direction of the photon momentum. The left part of the figure assumes
final kinetic energies $T_{p,1}=156\,{\rm MeV}$ and $T_{p,2}=33\,{\rm
MeV}$ of the two protons while in the right part the final kinetic
energies are $T_{p,1}=116\,{\rm MeV}$ and $T_{p,2}=73\,{\rm MeV}$.
The photon energy was chosen to be $\omega=215\,{\rm MeV}$ in all cases.
Together with the total structure functions (solid line) the
contributions arising from correlations (dashed line) and IC (dot-dashed
line) are shown. Please note the different scales.}
\end{figure}

As a first example, we would like to consider the case of ($e,e'pp$)
reactions in the so-called 'super parallel' kinematical situation where
the momentum $\vec{p}_1{'}$ of one knocked-out proton has the same
direction as the photon momentum.  The momentum of the second proton
$\vec{p}_2{'}$ is opposite to this direction.  The photon energy is fixed
at $\omega=215\,{\rm MeV}$. Results are shown in Figure \ref{fig5}. The
left section of this figure displays results selecting an asymmetric
distribution of the photon energy to the two protons. The one with final
momentum parallel to $\vec q$ is chosen to have a final kinetic energy of
$T_{p,1}=156\,{\rm MeV}$ while the second one with momentum antiparallel
to $\vec q$ has a final energy of $T_{p,2}=33\,{\rm MeV}$. For comparison,
we also present the example in which the photon energy is distributed in a
more symmetric way with $T_{p,1}=116\,{\rm MeV}$ and $T_{p,2}=73\,{\rm
MeV}$ in the right part of Figure \ref{fig5}. The upper part of this
figure exhibits results for the longitudinal structure function $W_L$
while the transverse structure function $W_T$ is displayed in the lower
parts of the figure.

As it has been discussed in the previous section, the MEC originating from
$\pi$ exchange do not contribute to the photon induced two-proton knockout
process. The isobar currents (IC) only affect the transverse structure
function $W_T$. Therefore, the results for the longitudinal structure
functions $W_L$ displayed in the upper part of Figure \ref{fig5} are
solely due to the effects of correlations. We can see that this
correlation contribution $W_L$ is significantly larger in the case of
asymmetric energy distribution (upper left part of the figure) than in the
case of more symmetric energy distribution. Note the different scales on
these parts of the figures. The correlation contribution `prefers'
processes in which the momentum and the energy of the virtual photon is
absorbed by one proton which is knocked-out parallel to the momentum $\vec
q$, while the second proton receives only a smaller fraction.

The situation is rather similar for the transverse structure function
$W_T$. Also here, the cross section due to correlations (dashed line) is
significantly larger in the case of asymmetric energy distribution. The
contribution of the IC to the structure function (dashed-dotted lines) is
similar in magnitude for the two cases considered. The correlation and
isobar contributions to the total structure function $W_T$ (solid lines)
are of similar importance for the more symmetric distribution of energy,
while the correlation contribution is dominating the cross section in the
asymmetric case, in particular at larger photon momenta.

\begin{figure}[h]
\hspace{1cm}\epsfig{file=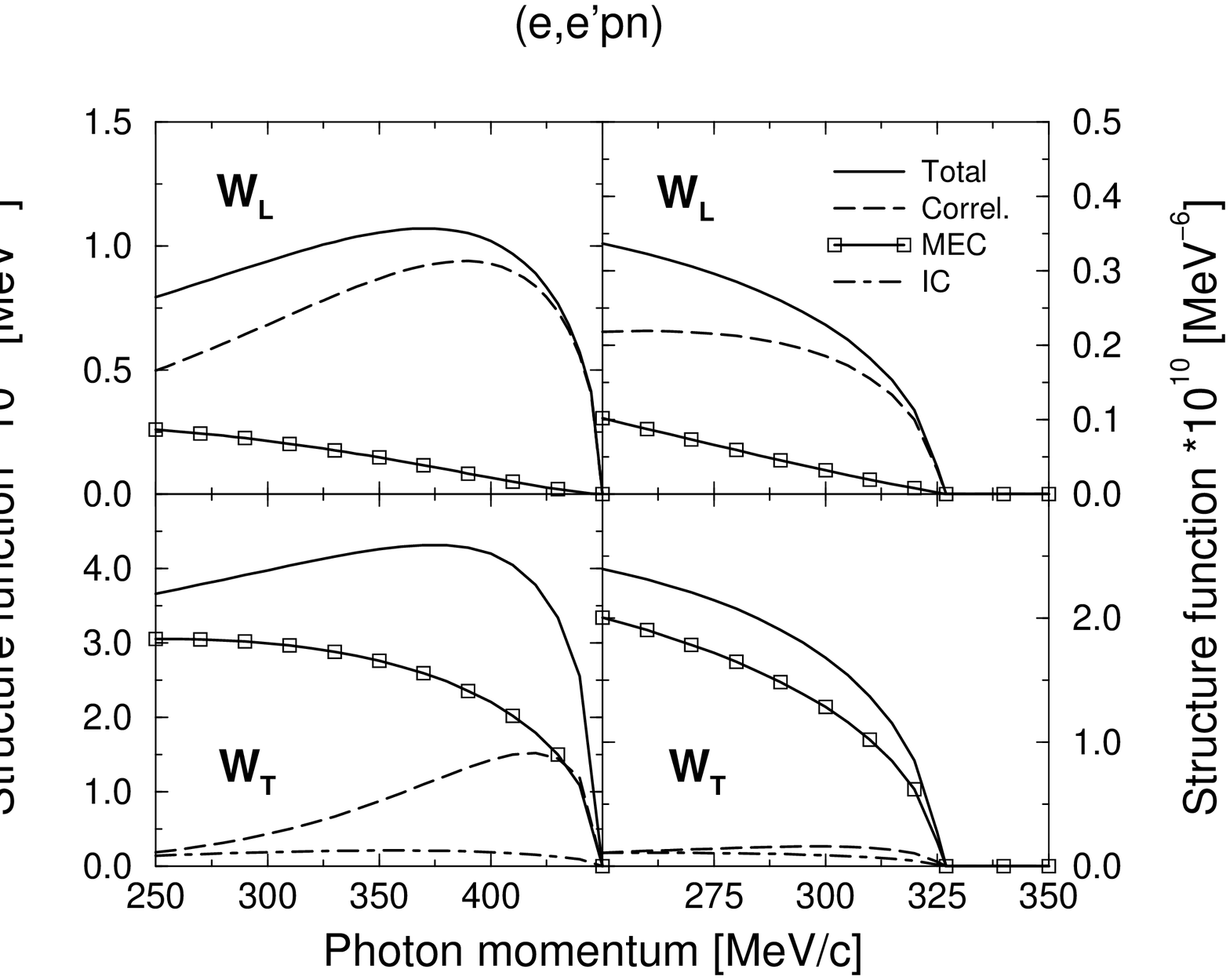,scale=0.7}
\vspace{.5cm}
\caption{\label{fig6}
Longitudinal (upper part) and transverse structure functions (lower part)
for the knockout of a proton-neutron pair in a 'super parallel'
kinematical situation with angles $\theta_{p}'=0^{\rm o}$ and
$\theta_{n}'=180^{\rm o}$ of the proton and the neutron with respect to
the direction of the photon momentum. The left part of the figure assumes
final kinetic energies $T_{p}=156\,{\rm MeV}$ and $T_{n}=33\,{\rm MeV}$ of
the two protons while in the right part the final kinetic energies are
$T_{p}=116\,{\rm MeV}$ and $T_{n}=73\,{\rm MeV}$. The photon energy was
chosen to be $\omega=215\,{\rm MeV}$ in all cases. Together with the total
structure functions (solid line) the contributions arising from
correlations (dashed line), MEC (solid line with squares) and IC
(dot-dashed line) are shown. Please note the different scales.}
\end{figure}

As a next example, we discuss the ($e,e'pn$) reaction under the same
kinematical conditions. Here, we consider the case with the proton emitted
parallel to the photon momentum and the neutron momentum antiparallel to
$\vec q$. In this case, we have to take into account the effects of MEC as
well (solid line with squares).  Comparing the knockout of a
proton-neutron pair to the ejection of a pair of protons, one observes
that the structure functions for the $(e,e'pn)$ reaction are roughly 10
times bigger than the structure functions for the $(e,e'pp)$ reactions
(see Figure \ref{fig6}). This fraction is slightly larger for transverse
photon polarisation than for longitudinally polarized virtual
photons. This enhancement of the structure functions is, of course, partly
due to the MEC. We find, however, that the contribution of correlations to
the structure functions alone are enhanced by roughly a factor of 7
comparing ($e,e'pn$) with $(e,e'pp)$. This demonstrates the importance of
tensor correlations which are more effective in the $pn$ than in the $pp$
interaction. We would like to recall that we have used the potential BONN A for
the NN interaction. This interaction exhibits a rather weak tensor component as
compared to other models for a realistic NN interaction. A measure for the
strength of this tensor component is given by the D-state probability in the
deuteron wavefunction. The BONN A potential yields a D-state probability of only
4.38 percent, while other meson exchange potentials predict D-state
probabilities up to 6 percent\cite{machleidt}. 
Again, we notice that the correlation effect is stronger when
the photon energy and momentum is distributed in a rather asymmetric way
to the knocked-out nucleons (left column of Figure \ref{fig6}).

In the 'super parallel' kinematical setup, the MEC contribution is of
particular importance in the transverse structure function (lower sections
of Figure \ref{fig6}). It dominates the corresponding cross section in the
case of the more symmetric energy distribution. In the asymmetric case,
however, the ratio of MEC to correlation contribution depends in quite a
sensitive way on the photon momenta and is in the average twice as large
as the correlation contribution for the cases considered. The IC
contribution is small in all these cases since the photon energy
considered ($\omega$ = 215 MeV) is below the threshold of the $\Delta$.

\begin{figure}[h]
\hspace{1cm}\epsfig{file=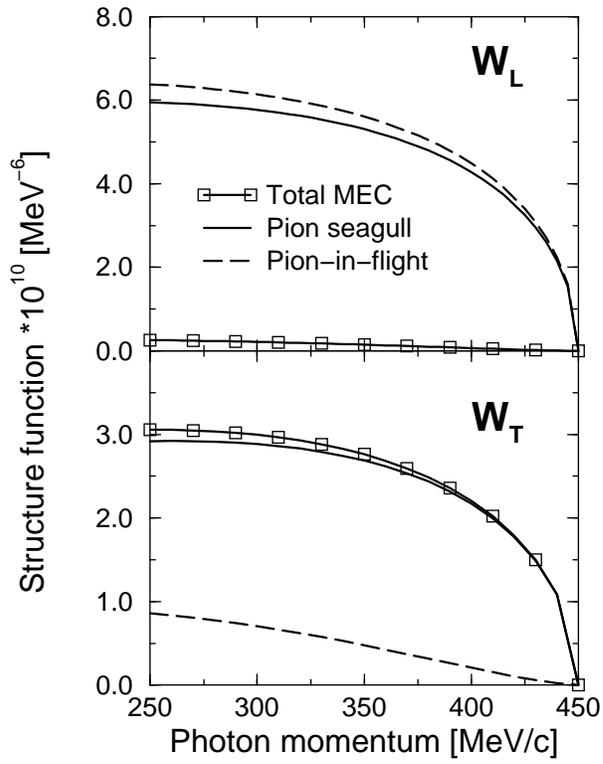,scale=0.7}
\vspace{.5cm}
\caption{\label{fig7}
MEC contribution for the asymmetric kinematical situation of Figure
\ref{fig6} (right part). The total MEC contribution (solid line with
squares) consists of the $\pi$-seagull (solid line) and the
$\pi$-in-flight contribution (dashed line).}
\end{figure}

In order to investigate the dependence of the MEC contribution on the kind
of structure function, we display in Figure \ref{fig7} the contribution of
the $\pi$-seagull terms [diagrams (a) and (b) of Fig.~\ref{fig3}] and the
$\pi$-in-flight term [Fig.~\ref{fig3} (c)] separately for the asymmetric
kinematic discussed in Figure \ref{fig6}. These contributions are of
similar importance in the longitudinal structure function but cancel each
other almost completely. Such a strong cancellation does not occur in the
transverse channel in which the $\pi$-in-flight term dominates. Therefore,
one may consider the longitudinal structure function for ($e,e'pn$)
reactions in 'super parallel' kinematics as an ideal setup to study
effects of correlations.

\begin{figure}[h]
\hspace{1cm}\epsfig{file=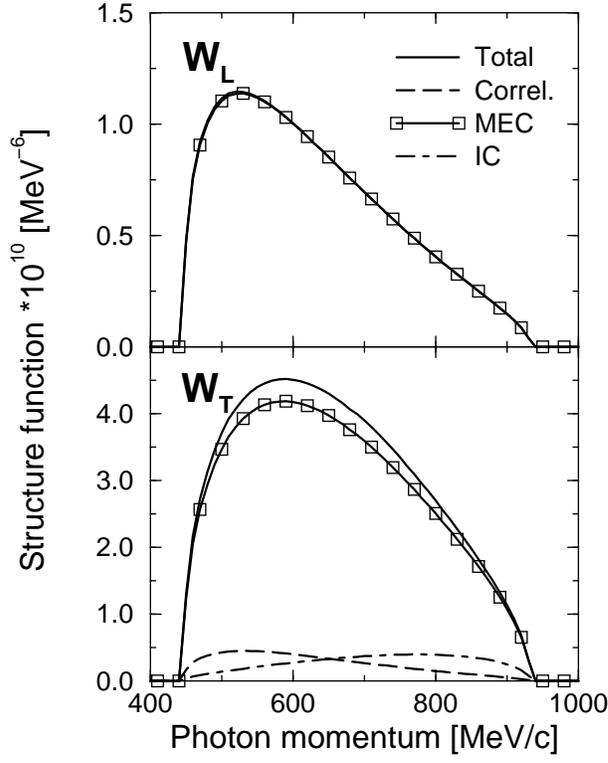,scale=0.7}
\vspace{.5cm}
\caption{
Longitudinal (above) and transverse structure functions (below) for the
knockout of a proton-neutron pair. The angles of the outgoing nucleons are
$\theta_{p}'=\theta_{n}'=30^{\rm o}$ while the final
kinetic energies are $T_{p}=T_{n}=70\,{\rm MeV}$. The photon energy was
chosen to be $\omega=230\,{\rm MeV}$. The total structure functions (solid 
line) consist of the one-body current (dashed line), the MEC
contribution (solid line with squares) and the IC contribution
(dot-dashed line).} 
\label{fig8}
\end{figure}

A large contribution from MEC is observed in other kinematical setups. As
an example, we show in Figure \ref{fig8} the longitudinal and transverse
structure functions depending on the photon momentum $|\vec{q}\,|$ for a
symmetric splitting of the kinetic energies and the angles with respect to
the direction of the momentum transfer of the outgoing particles
($T_p=T_n=70\,{\rm MeV}$ and $\theta_p'=\theta_n'=30^{\rm o}$).  The
photon energy was chosen to be $\omega=230\,{\rm MeV}$.  In this
framework, the influence of the single-particle current (correlations) is
almost negligible compared to the dominating MEC contribution. At higher
photon momenta, the isobaric current contribution gets even more important
than the correlation contribution.
  
\begin{figure}[h]
\hspace{1cm}\epsfig{file=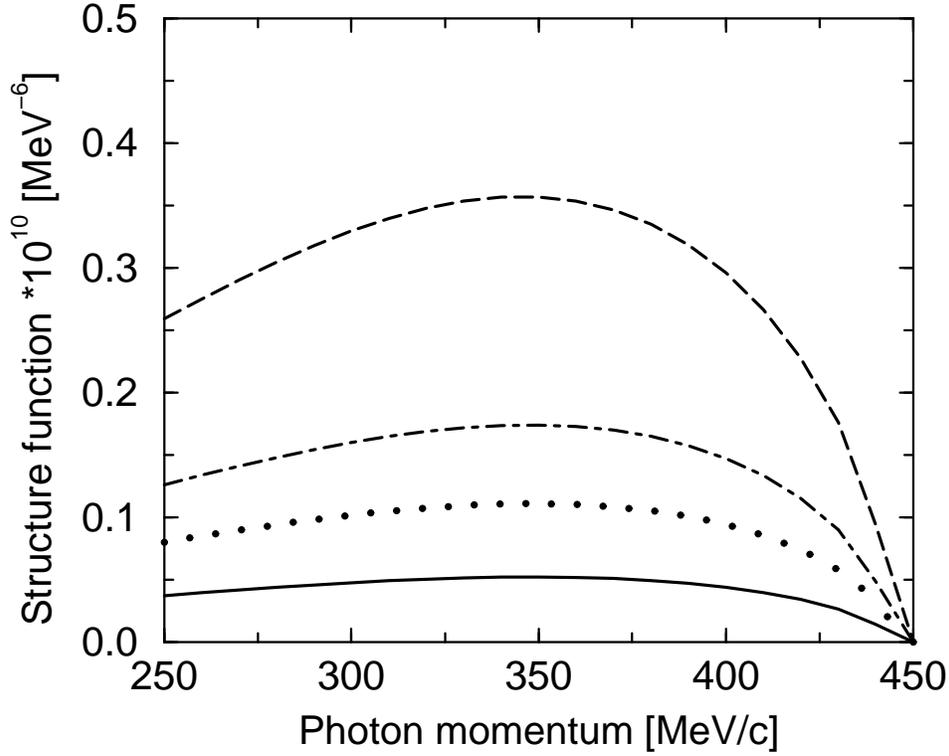,scale=0.7}
\vspace{.5cm}
\caption{\label{fig9}
The influence of the choice of the $\Delta$ propagator on the magnitude
of the isobaric current contribution. Referring to the kinematical
situation of Figure \ref{fig6}, matrix elements squared for different
$\Delta$ propagators are shown. The solid line represents the propagator
used in our calculations (c.f. eq. \ref{delprop}). A propagator of the
form (\ref{propag1})
for both resonant and non-resonant cases yields the dashed line, whereas
the reduction to the static propagator (\ref{propag2}) for the
non-resonant case leads to the dot-dashed line. Finally, a propagator of
the form suggested in \protect\cite{ryck}
is reflected by the dotted curve. For further
details, please refer to section II.C and the numerical results section.
}
\end{figure}

Finally, we would like to add a few remarks on the evaluation of the
isobar contribution to the two-body current.  As an example, we consider
in Figure \ref{fig9} the IC contribution to the transverse structure
function of ($e,e'pn$) using the kinematical setup of Figure \ref{fig6}
(left part). The IC contribution is rather sensitive to the approximation
employed for the $\Delta$ propagator \cite{disc1,disc2}. The correct
evaluation of the propagator (\ref{delprop}) for the resonant [Figure
\ref{fig4} (a) and (b)] and non-resonant [Figure \ref{fig4} (c) and (d)]
terms, as discussed above, leads to an IC contribution (solid line) which
is significantly smaller than the one obtained if a propagator of the form
(\ref{propag1}) is used for the resonant as well as the non-resonant case
(dashed line). The static propagator (\ref{propag2}) replacing the energy
denominators for the propagators in all four terms of the Figure
\ref{fig4} by the $\Delta$ nucleon mass difference leads to a structure
function (dashed-dotted line) which is roughly a factor of three larger
than the exact result and also the propagator suggested in \cite{ryck}
(dotted line)  overestimates the correct result by roughly a factor of
two.

\section{Summary and Conclusions}

The contributions originating from nucleon-nucleon correlations, meson-exchange
currents (MEC) and isobar currents (IC) to the cross section for photon induced
two-nucleon knock-out reactions ($e,e'2N$) have been investigated by evaluating
the corresponding matrix elements in  a consistent way for a system of infinite
nuclear matter. The relative importance of these various contributions is quite
sensitive to the kinematical setup of such triple coincidence experiments. The
so called super-parallel kinematic turns out to be very appropriate for the
investigation of NN correlation effects. This is particularly true for reactions
in which momentum and energy of the absorbed photon are transfered predominantly
on one of the nucleons knocked-out: Correlations seem  not to be very
`efficient' to transfer large amounts of energy to the second nucleon.

The ($e,e'pp$) reactions are favourable for the study of NN correlations
because the competing two-body current contributions are suppressed: The
MEC from $\pi$-exchange which dominate ($e,e'pn$) reactions don't
contribute in this case and IC only show up in the transverse structure
function. However, it is very important to investigate ($e,e'pn$)
reactions as well. The $pn$ correlations are quite different from the $pp$
correlations, the structure functions originating from correlation effects
are by a factor of 5 to 10 larger for the $pn$ knock-out than for
($e,e'pp$).

The MEC contributions which are dominating in many kinematical setups, are
highly suppressed in the 'super parallel' kinematics in which the proton
is observed in the direction of the photon and the neutron is detected
anti-parallel to the momentum of the photon. The suppression of the MEC is
due to a strong cancellation between the $\pi$-seagull and $\pi$-in-flight
contribution. The correlation effects should show up in particular in
the longitudinal structure function in this kinematical setup.

The present study includes the effects of a final state interaction
between the knocked-out nucleons and the residual interaction on the level
of a mean-field approximation. One should also consider, however, the
contribution to the cross section which originates from the direct
interaction of the two nucleons after the absorption of the photon. This
can be done in the scheme outlined in this manuscript. The importance of the
final state interaction has been demonstrated e.g.~in the Monte-Carlo 
simulations of ref.\cite{gil}. The consistent calculation
scheme for correlations, two-body currents and final state interaction
should also be applied
directly to the evaluation of cross section for specific target nuclei.

The authors wish to thank Peter Grabmayr for fruitful discussions.
This research project has partially been supported by the SFB 382 of the
Deutsche Forschungsgemeinschaft and the DFG-Graduiertenkolleg GRK 132.

\vfil

\medskip

\newpage

\medskip\vfil

\newpage

\newpage

\newpage

\newpage

\newpage

\end{document}